\documentclass[aps, prd, twocolumn, showpacs ]{revtex4}

\usepackage{amsmath,amsfonts,amscd,amssymb,mathrsfs,epsf}
\usepackage[small]{caption}

\newcommand{\M}{\mathcal{M}}

\newcommand{\vep}{\varepsilon}

\begin{document}

 \title{Casimir Interaction Between Two Concentric Cylinders at Nonzero Temperature}
 \author{L. P. Teo}\email{ LeePeng.Teo@nottingham.edu.my}
\address{Department of Applied Mathematics, Faculty of Engineering, University of Nottingham Malaysia Campus, Jalan Broga, 43500, Semenyih, Selangor Darul Ehsan, Malysia.}

\begin{abstract}
We study the finite temperature Casimir interaction between two concentric cylinders. When the separation between the cylinders is much smaller than the radii of the cylinders, the   asymptotic expansions of the Casimir interaction    are derived. Both the low temperature and the high temperature regions are considered. The leading terms are found to agree with the proximity force approximations.  The low temperature leading term of the temperature correction is also computed and it is found to be independent of the boundary conditions imposed on the larger cylinder.
\end{abstract}

\maketitle

\section{Introduction}
In recent years, there is an intensive interest in studying the Casimir interactions between two objects due to the success in experimental verification of the Casimir effect \cite{5,6,7,8,9,10,11} and the advent in the field theoretical method for computing the Casimir interactions \cite{12,14,15,16,17,20,21,22,24,25,26,27,28,29}. The Casimir interactions between spheres, between cylinders, between a plane and a sphere and between a plane and  a cylinder, have been extensively studied. However, most of the studies only considered the zero temperature interactions. Nevertheless, the finite temperature corrections have started to gain more attention  recently \cite{30,31,32,33,34,35}. In this letter, we study the finite temperature Casimir interaction between two concentric cylinders.

The zero temperature Casimir interaction between two concentric cylinders has been considered in \cite{18,19,1,4,2,3}. Here we consider the finite temperature correction. Of particular interest is the asymptotic behaviors of the Casimir interaction when the separation between the spheres is small. In \cite{4}, the asymptotic behavior at zero temperature has been obtained for two concentric perfectly conducting cylinders. Here we would consider both scalar fields and electromagnetic fields with different combinations of boundary conditions, which include Dirichlet-Dirichlet (DD), Neumann-Neumann (NN), Dirichlet-Neumann (DN -- Dirichlet on the smaller cylinder and Neumann on the larger cylinder) and Neumann-Dirichlet (ND) for scalar fields; and perfectly conducting - perfectly  conducting (PC-PC) and perfectly conducting - infinitely permeable (PC-IP) for electromagnetic fields. In fact, for cylinders, the results for electromagnetic fields can be easily derived from the results for scalar fields.

  \begin{figure}[h]
\epsfxsize=0.8\linewidth \epsffile{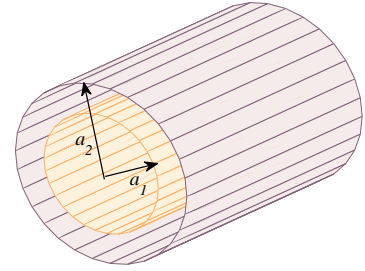} \caption{\label{f1} Two concentric cylinders with radius $a_1<a_2$. }\end{figure}
Consider two   concentric cylinders with radii $a_1<a_2$ (see Fig. \ref{f1}). Let $d=a_2-a_1$ be the separation between the cylinders.
At zero temperature, the Casimir interaction energy  between the two cylinders is given by \cite{1,2,3,4}:
\begin{equation}\label{eq6_9_8}\begin{split}
E_{\text{Cas}}^{T=0}=&\frac{L}{2\pi^2}\sum_{n=-\infty}^{\infty}\int_0^{\infty}\int_0^{\infty} \ln \left(1-M_n\left(\sqrt{\xi^2+k^2}\right)\right)d\xi dk\\=&\frac{L}{2\pi}\sum_{n=0}^{\infty}\!'\int_0^{\infty}\xi \ln \left(1-M_n(\xi)\right)d\xi,
\end{split}\end{equation}where $L$ is the length of the cylinders and
\begin{equation*}
\begin{split}
M_n(\xi)=&\frac{\mathscr{Z}_n^1(\xi)}{\mathscr{Z}_n^2(\xi)},
\end{split}
\end{equation*}with
\begin{equation*}
\begin{split}
\mathscr{Z}_n^i(\xi)=\frac{I_n(a_i\xi)}{K_n(a_i\xi)}\quad \text{or}\quad \mathscr{Z}_n^i(\xi)=\frac{I_n'(a_i\xi)}{K_n'(a_i\xi)}
\end{split}
\end{equation*}depending on whether Dirichlet or Neumann  condition is imposed on the cylinder with radius $a_i$. Here $I_n(z)$ and $K_n(z)$ are modified Bessel functions.

Using Matsubara representation, the finite temperature Casimir free energy is  given by
\begin{equation}\label{eq6_8_1}\begin{split}
E_{\text{Cas}} =&\frac{TL}{\pi}\sum_{n=-\infty}^{\infty}\sum_{l=0}^{\infty}\!'\int_0^{\infty} \ln \left(1-M_n\left(\sqrt{\xi_l^2+k^2}\right)\right) dk\end{split}\end{equation}\begin{equation*}\begin{split}=&\frac{2TL}{\pi}\sum_{n=0}^{\infty}\!'\sum_{l=0}^{\infty}\!'\int_{\xi_l}^{\infty}\frac{\xi}{\sqrt{\xi^2-\xi_l^2}} \ln \left(1-M_n(\xi)\right)d\xi,
\end{split}\end{equation*}where $\xi_l=2\pi l T$ are the Matsubara frequencies. Using Poisson resummation formula, \eqref{eq6_8_1} can be rewritten as
\begin{equation}\label{eq6_8_2}
E_{\text{Cas}}=E_{\text{Cas}}^{T=0}+\frac{L}{\pi}\sum_{n=0}^{\infty}\!'\sum_{l=1}^{\infty} \int_0^{\infty}\xi J_0\left(\frac{l\xi}{T}\right)  \ln \left(1-M_n(\xi)\right)d\xi,
\end{equation}where $J_{n}(z)$ is the Bessel function of first kind. Eq. \eqref{eq6_8_1} is the so-called high temperature expansion of the Casimir free energy and \eqref{eq6_8_2} is the so-called low temperature expansion. The latter shows manifestly the thermal correction to the Casimir free energy, which is given by
\begin{equation} \Delta_TE_{\text{Cas}}=\frac{L}{\pi}\sum_{n=0}^{\infty}\!'\sum_{l=1}^{\infty} \int_0^{\infty}\xi J_0\left(\frac{l\xi}{T}\right)  \ln \left(1-M_n(\xi)\right)d\xi.\end{equation}

For electromagnetic field with perfectly conducting   (or infinitely permeable) condition on both cylinders, the Casimir free energy  is the sum of the Casimir free energies for DD and NN boundary conditions. If one cylinder is perfectly conducting and one is infinitely permeable, then the Casimir free energy is the sum of the Casimir free energies for DN and ND boundary conditions.

In the following, we derive the asymptotic expansions of the Casimir free energy when the separation between the cylinders is small compared to the radii of the cylinders. We consider the low temperature region where $dT\ll a_2T\ll 1$ and the high temperature region where $1\ll dT\ll a_2T$.
In the low temperature region, the dominating term of the Casimir free energy is the zero temperature term \eqref{eq6_9_8}, whereas
in the high temperature region, the dominating term  is the classical term:
  \begin{equation}\label{eq6_9_7}
E_{\text{Cas}}^{\text{cl}} = \frac{ TL}{\pi}\sum_{n=0}^{\infty}\!' \int_{0}^{\infty}  \ln \left(1-M_n(\xi)\right)d\xi,
\end{equation}
whose Matsubara frequency is zero.
Define the dimensionless parameter $$\vep=\frac{a_2-a_1}{a_1}=\frac{d}{a_1}.$$   Making a change of variables $\omega=a_1\xi$ in \eqref{eq6_9_8} and \eqref{eq6_9_7}, we have respectively
\begin{equation}\label{eq6_8_3}\begin{split}
E_{\text{Cas}}^{T=0} =& \frac{  L}{2\pi a_1^2}\sum_{n=0}^{\infty}\!' \int_{0}^{\infty} \omega\ln \left(1-A_n(\omega)\right)d\omega,\\
E_{\text{Cas}}^{\text{cl}} =& \frac{ TL}{\pi a_1}\sum_{n=0}^{\infty}\!' \int_{0}^{\infty} \ln \left(1-A_n(\omega)\right)d\omega,\end{split}
\end{equation}
where
$$A_n(\omega)=\frac{\mathscr{H}_n^1(\omega)}{\mathscr{H}_n^2(\omega(1+\vep))},$$$$\mathscr{H}_n^i(\omega)=\frac{I_n(\omega)}{K_n(\omega)}\quad\text{or}\quad
\mathscr{H}_n^i(\omega)=\frac{I_n'(\omega)}{K_n'(\omega)}$$
depending on whether Dirichler or Neumann  condition is imposed on the cylinder with radius $a_i$.
To unify the treatment for the zero temperature Casimir energy and the classical term of the Casimir free energy, define
\begin{equation}
\mathcal{E}_{\chi}=\sum_{n=0}^{\infty}\!' \int_{0}^{\infty} \omega^{\chi}\ln \left(1-A_n(\omega)\right)d\omega,
\end{equation}where $\chi=0$ or $1$.
Separating the terms with $n=0$ and the terms with $n\geq 1$, we  have $\mathcal{E}_{\chi}=\mathcal{E}_{\chi,0}+\mathcal{E}_{\chi,r}$. Expanding the logarithm gives
\begin{equation}\label{eq6_8_4}
\begin{split}
\mathcal{E}_{\chi, 0}=& - \frac{1}{2 } \sum_{s=1}^{\infty}\frac{1}{s}\int_{0}^{\infty} \omega^{\chi} A_0(\omega)^sd\omega,\\
 \mathcal{E}_{\chi,r}=&  - \sum_{n=1}^{\infty}n^{\chi+1}\sum_{s=1}^{\infty}\frac{1}{s}\int_{0}^{\infty} \omega^{\chi}A_n(n\omega)^sd\omega.
\end{split}\end{equation}
For the term with $n=0$, we need to use asymptotic expansions of $I_0(z)$ and $K_0(z)$ when $z\rightarrow \infty$, which give
\begin{equation*}
\begin{split}
\frac{I_0(\omega)}{K_0(\omega)}\sim &\frac{1}{\pi}\exp\left(2\omega+\ldots\right),\\
\frac{I_0'(\omega)}{K_0'(\omega)}\sim &-\frac{1}{\pi}\exp\left(2\omega+\ldots\right).
\end{split}
\end{equation*}
For the term with $n\neq 0$, Debye asymptotic expansions for Bessel functions \cite{36} show  that
\begin{equation*}
\begin{split}
\frac{I_n(n\omega)}{K_n(n\omega)}=\frac{1}{\pi}\exp\left(2n\eta(\omega)+ \frac{2D_{ 1}(t(\omega))}{n }+\ldots\right),\\
\frac{I_n'(n\omega)}{K_n'(n\omega)}=-\frac{1}{\pi}\exp\left(2n\eta(\omega)+ \frac{2M_{ 1}(t(\omega))}{n }+\ldots\right),
\end{split}
\end{equation*}where
\begin{equation*}\begin{split}\eta(z)=&\sqrt{1+z^2}+\log\frac{z}{1+\sqrt{1+z^2}},\quad t(z)=\frac{1}{\sqrt{1+z^2}},\\
 D_1(t)=&\frac{t}{8}-\frac{5t^3}{24},\hspace{1cm} M_1(t)=-\frac{3t}{8}+\frac{7t^3}{24}.\end{split}\end{equation*}
In the following, we consider the case of homogeneous boundary conditions (DD or NN) and mixed boundary conditions (DN or ND) separately.

For homogeneous boundary conditions,
\begin{equation*}
\begin{split}
A_0(\omega)\sim  \exp\left(-2\vep\omega+\ldots\right).
\end{split}
\end{equation*}  Therefore,
\begin{equation}\label{eq6_9_5}
\begin{split}
\mathcal{E}_{\chi, 0}\sim &- \frac{1}{2}\sum_{s=1}^{\infty}\frac{1}{s}\int_{0}^{\infty} \omega^{\chi}\exp\left(-2s\vep\omega \right)d\omega
\\=&\left\{\begin{aligned}&-\frac{\pi^2}{24\vep}+o(1),\quad &\chi=0\\
&-\frac{\zeta_R(3)}{8\vep^2}+o(\vep^{-1}),\quad &\chi=1\end{aligned}\right..
\end{split}
\end{equation}
On the other hand,
\begin{equation*}
\begin{split}
A_n(n\omega)\sim &\exp\Biggl(-2n\left[\eta([1+\vep]\omega)-\eta(\omega)\right]\\& \hspace{0.5cm}-2\frac{P_1(t([1+\vep]\omega))-P_1(t(\omega))}{n}\Biggr),
\end{split}
\end{equation*}where
$$P_1(t)=\lambda_0 t+\lambda_1 t^3=\left\{\begin{aligned}&D_1(t),\;&\text{for DD b.c.}\\
&M_1(t),\;&\text{for NN b.c.}\end{aligned}\right..$$
Applying the inverse Mellin transform formula \begin{equation}\label{eq6_1_4}e^{-v}=\frac{1}{2\pi i}\int_{c-i\infty}^{c+i\infty}   \Gamma(z) v^{-z}dz\end{equation}gives
\begin{equation}
\begin{split}
\mathcal{E}_{\chi,r}=& -\sum_{n=1}^{\infty}n^{\chi+1} \frac{1}{2\pi i}\int_{c-i\infty}^{c+i\infty} \Gamma(z) \zeta_R(z+1)2^{-z}n^{-z}\\&\hspace{1cm}\times\int_{0}^{\infty}\omega^{\chi} \left(\eta([1+\vep]\omega)-\eta(\omega)\right)^{-z}\\&\times\left(1+\frac{1}{n^2}\frac{P_1(t([1+\vep]\omega))-P_1(t(\omega))}{\eta([1+\vep]\omega)-\eta(\omega)}\right)^{-z}d\omega dz\\
=&-  \frac{1}{2\pi i}\int_{c-i\infty}^{c+i\infty} \Gamma(z) \zeta_R(z+1)2^{-z}\vep^{-z} \\&\times \Bigl(\zeta_R(z-\chi-1)\mathcal{A}_{\chi} (z)
-z\zeta_R(z-\chi+1)\mathcal{B}_{\chi}(z)\Bigr) dz,
\end{split}
\end{equation}where
\begin{equation*}
\begin{split}
\mathcal{A}_{\chi}(z)=& \int_{0}^{\infty}\omega^{\chi-z}\eta'(\omega)^{-z}\left(1-z\frac{\vep \omega }{2}\frac{\eta^{\prime\prime}(\omega)}{\eta'(\omega)}\right.\\&\left.+\vep^2\left[- \frac{ z \omega^2}{6}\frac{\eta^{\prime\prime\prime}(\omega)}{\eta'(\omega)}+ \frac{  z(z+1)\omega^2 }{8}\frac{\eta^{\prime\prime}(\omega)^2}{\eta'(\omega)^2}\right] \right)d\omega,\\
\mathcal{B}_{\chi}(z)=& \int_{0}^{\infty}\omega^{\chi-z}\eta'(\omega)^{-z}\frac{P_1'(t(\omega))}{\eta'(\omega)}t'(\omega)  d\omega.
\end{split}\end{equation*}
Straightforward computations give\begin{equation*}\begin{split}
 \mathcal{A}_{\chi}(z)=&\frac{\Gamma\left(\frac{\chi+1}{2}\right)}{2}\frac{\Gamma\left(\frac{z-\chi-1}{2}\right)}{\Gamma\left(\frac{z}{2}\right)}\Biggl\{ 1+\vep \frac{z-\chi-1}{2}\\&+\vep^2\frac{(z-\chi-1)(3z^2-2z-17-7\chi-3\chi z)}{24(z+2)}\Biggr\},\\
 \mathcal{B}_{\chi}(z)=&\frac{\Gamma\left(\frac{\chi+1}{2}\right)}{2}\frac{\Gamma\left(\frac{z-\chi+1}{2}\right)}{\Gamma\left(\frac{z+2}{2}\right)}\Biggl\{
 -\lambda_0
 +(\lambda_0-3\lambda_1)\frac{  z-\chi+1 }{z+2}\\&+3\lambda_1\frac{ \left( z-\chi+1 \right)\left(z-\chi+3\right)}
 {(z+2)(z+4)} \Biggr\}.
\end{split}
\end{equation*}
Therefore,
\begin{equation*}
\begin{split}
 \mathcal{E}_{0,r}=
&-\frac{\pi}{8\vep^2}\zeta_R(3)\left(1+\frac{\vep}{2}\right)+\frac{\pi^2}{24\vep}\\&+\frac{\pi}{16}\left(4\lambda_0+3\lambda_1\right)\ln\vep+O(1), \\
 \mathcal{E}_{1,r}=&-\frac{\pi^4}{360\vep^3}\left(1+\frac{\vep}{2}-\frac{\vep^2}{10}\right)+\frac{\zeta_R(3)}{8\vep^2}\\&-\frac{\pi^2}{6\vep}\left(\frac{\lambda_0}{3}
+\frac{\lambda_1}{5}\right)+o(\vep^{-1}).
\end{split}
\end{equation*}
Combining with \eqref{eq6_9_5}, we find that
\begin{equation*}
\begin{split}
\mathcal{E}_{0 }=
&-\frac{\pi}{8\vep^2}\zeta_R(3)\left(1+\frac{\vep}{2} - \left[2\lambda_0+\frac{3}{2}\lambda_1\right]\vep^2\ln\vep+O(\vep^2)\right),\\
\mathcal{E}_{1}=  &-\frac{\pi^4}{360\vep^3}\left(1+\frac{\vep}{2}-\frac{\vep^2}{10}+\frac{\vep^2}{\pi^2} \left[20 \lambda_0 +12\lambda_1 \right]+o(\vep^2)\right).\end{split}
\end{equation*}
From these, we find that the   asymptotic expansions of the zero temperature Casimir energies for DD and NN boundary conditions are given  respectively by
\begin{equation}
\begin{split}
E_{\text{Cas}}^{\text{DD},T=0}\sim &-\frac{\pi^3L}{720a_1^2\vep^3}\left(1+\frac{\vep}{2}-\frac{\vep^2}{10}+o(\vep^{2}) \right),\\
E_{\text{Cas}}^{\text{NN},T=0}\sim &-\frac{\pi^3L}{720a_1^2\vep^3}\left(1+\frac{\vep}{2}-\vep^2\left[\frac{1}{10}+\frac{4}{\pi^2}\right]+o(\vep^{2}) \right).
\end{split}
\end{equation}The leading term $$-\frac{\pi^3L}{720a_1^2\vep^3}=-\frac{\pi^2 }{1440 d^3}\times 2\pi a_1 L$$ is  what one would expect from the proximity force approximation.
For the high temperature asymptotic behavior of the Casimir free energies,  we have
\begin{equation}
\begin{split}
 E_{\text{Cas}}^{\text{DD},\text{cl}}\sim &-\frac{LT}{8a_1\vep^2}\zeta_R(3)\left(1+\frac{\vep}{2} +\frac{1}{16}\vep^2\ln\vep+O(\vep^{2})\right),\\
  E_{\text{Cas}}^{\text{NN},\text{cl}}\sim &-\frac{LT}{8a_1\vep^2}\zeta_R(3)\left(1+\frac{\vep}{2} +\frac{5}{16}\vep^2\ln\vep+O(\vep^{2})\right).
\end{split}
\end{equation}  Again, the leading term
$$-\frac{LT}{8a_1\vep^2}\zeta_R(3)=-\frac{T}{16\pi d^2}\times 2\pi a_1L$$   agrees with the proximity force approximation.

For electromagnetic field with  perfect conductor condition on both the cylinders, we find that the    asymptotic expansion of the zero temperature Casimir energy is
\begin{equation*}
\begin{split}
E_{\text{Cas}}^{\text{PC-PC},T=0}\sim &-\frac{\pi^3L}{360a_1^2\vep^3}\left(1+\frac{\vep}{2}-\vep^2\left[\frac{1}{10}+\frac{2}{\pi^2}\right]+o(\vep^{2}) \right).
\end{split}
\end{equation*}This agrees with the result obtained in \cite{4}. On the other hand, the high temperature   asymptotic expansion is
\begin{equation*}
\begin{split}
 E_{\text{Cas}}^{\text{PC-PC},\text{cl}}\sim &-\frac{LT}{4a_1\vep^2}\zeta_R(3)\left(1+\frac{\vep}{2} +\frac{3}{16}\vep^2\ln\vep+O(\vep^{2})\right).
\end{split}
\end{equation*}

 For mixed boundary conditions,
\begin{equation*}
\begin{split}
A_0(\omega)\sim -\exp\left(-2\vep \omega+ \ldots\right).
\end{split}
\end{equation*}
 Therefore,
\begin{equation}\label{eq6_9_5}
\begin{split}
\mathcal{E}_{\chi, 0}\sim &- \frac{1}{2}\sum_{s=1}^{\infty}\frac{(-1)^s}{s}\int_{0}^{\infty} \omega^{\chi}\exp\left(-2s\vep\omega \right)d\omega
\\=&\left\{\begin{aligned}&\frac{\pi^2}{48\vep}+o(1),\quad &\chi=0\\
&\frac{3\zeta_R(3)}{32\vep^2}+o(\vep^{-1}),\quad &\chi=1\end{aligned}\right..
\end{split}
\end{equation}
On the other hand,
\begin{equation*}
\begin{split}
A_n(n\omega)\sim &-\exp\left(-2n\left[\eta([1+\vep]\omega)-\eta(\omega)\right]\right.\\&\left.\hspace{1cm}-2\frac{Q_1(t([1+\vep]\omega))-P_1(t(\omega))}{n}\right),
\end{split}
\end{equation*}where
$$\begin{pmatrix}P_1(t)\\Q_1(t)\end{pmatrix}=\begin{pmatrix}\lambda_{0} t+\lambda_1 t^3\\
\varpi_0t+\varpi_1t^3\end{pmatrix}=\left\{\begin{aligned}&\begin{pmatrix}D_1(t)\\M_1(t)\end{pmatrix},\;&\text{for DN b.c.}\\
&\begin{pmatrix}M_1(t)\\D_1(t)\end{pmatrix},\;&\text{for ND b.c.}\end{aligned}\right..$$
Applying the inverse Mellin transform formula \eqref{eq6_1_4} gives
\begin{equation}
\begin{split}
\mathcal{E}_{\chi,r}=& \frac{1}{2\pi i}\int_{c-i\infty}^{c+i\infty} \Gamma(z) (1-2^{-z})\zeta_R(z+1)2^{-z}\vep^{-z} \\& \times \Biggl(\zeta_R(z-\chi-1)\mathcal{A}_{\chi} (z)
-z\zeta_R(z-\chi+1)\mathcal{C}_{\chi}(z)\\&\hspace{1cm}+\frac{z(z+1)}{2\vep^2}\zeta_R(z-\chi+3)\mathcal{G}_{\chi} (z)\Biggr) dz,
\end{split}
\end{equation}
where $\mathcal{A}_{\chi}(z)$ is as before,
\begin{equation*}
\begin{split}
 \mathcal{C}_{\chi}(z)=&\int_0^{\infty}\omega^{\chi-z} \eta'(\omega) \left[\frac{Q_1'(t(\omega))t'(\omega)}{\eta'(\omega)} \right.\\& -(z+1)\frac{ Q_1(t(\omega))-P_1(t(\omega))}{2}\frac{\eta^{\prime\prime}(\omega)}{\eta^{\prime}(\omega)^2}\\&\left.+\frac{1}{\vep}\frac{ Q_1(t(\omega))-P_1(t(\omega))}{\omega\eta'(\omega)}\right]d\omega,\\
 =&\frac{\Gamma\left(\frac{\chi+1}{2}\right)}{2}\frac{\Gamma\left(\frac{z-\chi+1}{2}\right)}{\Gamma\left( \frac{z+2}{2}\right)}\\& \times\left( -\varpi_0 +\left(\varpi_0-3\varpi_1+\frac{z+1}{2}\kappa_0\right)\frac{z-\chi+1}{z+2}\right.\\&\left.+\left(3\varpi_1+\frac{z+1}{2}\kappa_1\right)\frac{(z-\chi+1)(z-\chi+3)}{(z+2)(z+4)}
\right.\\& \left.+\frac{1}{\vep}\left[\kappa_0+\kappa_1\frac{z-\chi+1}{z+2}\right]\right),\end{split}
\end{equation*}\begin{equation*}
\begin{split}
  \mathcal{G}_{\chi} (z)=&\int_0^{\infty}\omega^{\chi-z-2} \eta'(\omega)^{-z-2}\left( Q_1(t(\omega))-P_1(t(\omega))\right)^2d\omega\\
  =& \frac{\Gamma\left(\frac{\chi+1}{2}\right)}{2}\frac{\Gamma\left(\frac{z-\chi+3}{2}\right)}{\Gamma\left(\frac{z+4}{2}\right)}\left(\kappa_0^2+2\kappa_0\kappa_1 \frac{z-\chi+3}{z+4}\right.\\&\left.+\kappa_1^2\frac{(z-\chi+3)(z-\chi+5)}{(z+4)(z+6)}\right).
\end{split}
\end{equation*}
Here $\kappa_i=\varpi_i-\lambda_i$.
Therefore,
\begin{equation*}\begin{split}
\mathcal{E}_{0,r}=
& \frac{3\pi}{32\vep^2}\zeta_R(3)\left(1+\frac{\vep}{2}\right)-\frac{\pi^2}{48\vep}-\frac{\pi}{4\vep} \left(2\kappa_0+ \kappa_1\right)\ln 2\\&-\ln\vep
\left(\frac{\kappa_0}{2}+\frac{\pi}{4}\left[\kappa_0^2+\kappa_0\kappa_1+\frac{3}{8}\kappa_1^2\right]\right)+O(1),\\
\mathcal{E}_{1,r}=&\frac{7\pi^4}{2880\vep^3}\left(1+\frac{\vep}{2}-\frac{\vep^2}{10}\right)-\frac{\pi^2}{72\vep^2}(3\kappa_0+\kappa_1)\\&
-\frac{3\zeta_R(3)}{32\vep^2}+\frac{\pi^2}{24\vep}
\left(\frac{2\varpi_0-\kappa_0}{3}+\frac{2\varpi_1-\kappa_1}{5}\right)\\
& +\frac{\ln2}{2\vep}\kappa_0+\frac{1}{2\vep}\left(\kappa_0^2+\frac{2}{3}\kappa_0\kappa_1+\frac{1}{5}\kappa_1^2\right)+o(\vep^{-1}).\end{split}\end{equation*}
From these, we find that the   asymptotic behaviors of the zero temperature Casimir energies for DN and ND boundary conditions are given  respectively by
\begin{equation}
\begin{split}
E_{\text{Cas}}^{\text{DN},T=0}\sim &\frac{7\pi^3L}{5760a_1^2\vep^3}\left(1+\vep\left[\frac{1}{2}+\frac{40}{7\pi^2}\right]\right.\\&\left.+\vep^2\left[-\frac{1}{10}-\frac{8}{7\pi^2}-\frac{720}{7\pi^4}\ln2+\frac{192}{7\pi^4}\right]+o(\vep^{2}) \right),\\
E_{\text{Cas}}^{\text{ND},T=0}\sim &\frac{7\pi^3L}{5760a_1^2\vep^3}\left(1+\vep\left[\frac{1}{2}-\frac{40}{7\pi^2}\right]\right.\\&\left.+\vep^2\left[-\frac{1}{10}-\frac{8}{7\pi^2}+\frac{720}{7\pi^4}\ln2+\frac{192}{7\pi^4}\right]+o(\vep^{2}) \right).
\end{split}
\end{equation}The leading term $$\frac{7\pi^3L}{5760a_1^2\vep^3}=\frac{7\pi^2 }{11520 d^3}\times 2\pi a_1 L$$ agrees with the proximity force approximation.
For the high temperature asymptotic behaviors of the Casimir free energies,  we have
\begin{equation}
\begin{split}
 E_{\text{Cas}}^{\text{DN},\text{cl}}\sim &\frac{3LT}{32a_1\vep^2}\zeta_R(3)\left(1+\vep\left[\frac{1}{2}+\frac{4 }{3\zeta_R(3)}\ln 2\right] \right.\\&\left.+\left[\frac{8}{3\pi\zeta_R(3)}-\frac{1}{4\zeta_R(3)}\right]\vep^2\ln\vep+O(\vep^2)\right),\\
  E_{\text{Cas}}^{\text{ND},\text{cl}}\sim &\frac{3LT}{32a_1\vep^2}\zeta_R(3)\left(1+\vep\left[\frac{1}{2}-\frac{4 }{3\zeta_R(3)}\ln 2\right] \right.\\&\left.-\left[\frac{8}{3\pi\zeta_R(3)}+\frac{1}{4\zeta_R(3)}\right]\vep^2\ln\vep+O(\vep^2)\right).
\end{split}
\end{equation}  Again, the leading term
$$\frac{3LT}{32a_1\vep^2}\zeta_R(3)=\frac{3T}{64\pi d^2}\times 2\pi a_1L$$   agrees with the proximity force approximation.

For electromagnetic field with  perfect conductor condition on one cylinder and infinitely permeable condition on the other cylinder, we find that the    asymptotic expansion of the zero temperature Casimir energy is
\begin{equation}
\begin{split}
E_{\text{Cas}}^{\text{PC-IP},T=0}\sim &\frac{7\pi^3L}{2880a_1^2\vep^3}\left(1+ \frac{\vep}{2} \right.\\&\left.+\vep^2\left[-\frac{1}{10}-\frac{8}{7\pi^2} +\frac{192}{7\pi^4}\right]+o(\vep^{2}) \right);
\end{split}
\end{equation}  whereas the high temperature  asymptotic behavior of the Casimir free energy is
\begin{equation}
\begin{split}
 E_{\text{Cas}}^{\text{PC-IP},\text{cl}}\sim & \frac{3LT}{16a_1\vep^2}\zeta_R(3)\left(1+\frac{\vep}{2}-\frac{1}{4\zeta_R(3)}\vep^2\ln\vep  +O(\vep^{2})\right).
\end{split}
\end{equation}
From the results above, we see that the leading terms always agree with the proximity force approximations. The   correction terms are more complicated in the case of mixed boundary conditions, where the force is repulsive.

As we mentioned above, in the low temperature region, the Casimir free energy is dominated by the zero temperature term. However, it would be interesting to determine the order of magnitude of the temperature correction.
For this, the Abel-Plana summation formula  can be used to show that the temperature correction can be written as
\begin{equation}\label{eq5_12_1}
\begin{split}
\Delta_TE_{\text{Cas}}=& \frac{L}{ \pi^2}\sum_{n=0}^{\infty}\!' \int_0^{\infty}\int_0^{\infty}\frac{\xi d\xi}{ \sqrt{\xi^2+k^2}}\\&\times
\frac{i\bigl[   \ln (1-M_n(i\xi))-  \ln (1-M_n(-i\xi))\bigr]}{e^{ \frac{\sqrt{\xi^2+k^2}}{T}}-1}dk\\&+\text{exponentially decaying terms}.
\end{split}
\end{equation}From this, we see that in the low temperature region, the leading terms of the temperature correction  come from the term
\begin{equation}
\begin{split}
  \frac{L}{ \pi^2a_1^2}\sum_{n=0}^{\infty}\!' \int_0^{\infty}\int_0^{\infty}
i\mathscr{S}_n(\omega) \frac{1}{e^{ \frac{\sqrt{\omega^2+k^2}}{a_1T}}-1}\frac{\omega d\omega}{ \sqrt{\omega^2+k^2}}dk,
\end{split}
\end{equation}where
$$\mathscr{S}_n(\omega)=\ln (1-A_n(i\omega))-  \ln (1-A_n(-i\omega)).$$
Using
\begin{equation*}\begin{split}
&I_{n}(iz)=i^nJ_n(z),\hspace{1cm}K_n(iz)=-\frac{i\pi}{2}i^{-n}H_{n}^{(2)}(z),\\
&I_{n}(-iz)=i^{-n}J_n(z),\hspace{1cm}K_n(-iz)=\frac{i\pi}{2}i^{n}H_{n}^{(1)}(z),
\end{split}
\end{equation*}one can show that for DD or DN boundary conditions,
\begin{equation*}
\begin{split}
\mathscr{S}_n(\omega)=&
\ln\left( 1-i\frac{J_n (\omega)}{N_n (\omega)} \right)-\ln\left( 1+i\frac{J_n (\omega)}{N_n (\omega)} \right);
\end{split}
\end{equation*}whereas for ND or NN boundary conditions,
\begin{equation*}
\begin{split}
\mathscr{S}_n(\omega)=&
\ln\left( 1-i\frac{J_n' (\omega)}{N_n' (\omega)} \right)-\ln\left( 1+i\frac{J_n' (\omega)}{N_n' (\omega)} \right).\end{split}
\end{equation*}
From here, we see that in the low temperature region, the leading terms of the temperature correction do not depend on the boundary condition on the larger cylinder.
As $\omega\rightarrow 0$,
\begin{equation*}\begin{split}
\frac{J_n (\omega)}{N_n (\omega)}=O(\omega^{2n}),\quad\frac{J_n' (\omega)}{N_n' (\omega)}=O(\omega^{2n}),\quad n\geq 1;\end{split}\end{equation*}  and
\begin{equation*}\begin{split}
\frac{J_0(\omega)}{N_0(\omega)}\sim &\frac{\pi\left(1-\frac{\omega^2}{4}\right)+\ldots}{2\ln\frac{\omega}{2}\left(1-\frac{\omega^2}{4}\right)-2\psi(1)+\ldots }\sim\frac{\pi }{2\ln \omega }+\ldots,\\
\frac{J_0' (\omega)}{N_0' (\omega)}\sim &-\frac{\pi}{4}\omega^{2}+\ldots,\quad \frac{J_1' (\omega)}{N_1' (\omega)}\sim \frac{\pi}{4}\omega^{2}+\ldots.\end{split}\end{equation*}
From these, we find that if the smaller cylinder is Dirichlet, the leading contribution to the thermal correction comes from the term with $n=0$, which gives
\begin{equation}\label{eq6_30_1}
\begin{split}
\Delta_TE_{\text{Cas}}\sim  &\frac{L}{\pi a_1^2}\int_0^{\infty}\int_0^{\infty}
 \frac{1}{\ln\omega}\frac{1}{e^{ \frac{\sqrt{\omega^2+k^2}}{a_1T}}-1}\frac{\omega dkd\omega}{ \sqrt{\omega^2+k^2}}
 \\=&\frac{TL}{\pi a_1}\int_0^{\infty}\int_{\frac{\omega}{a_1T}}^{\infty}\frac{1}{\ln\omega} \frac{1}{e^{u}-1}\frac{\omega dud\omega}{\sqrt{(a_1Tu)^2-\omega^2}}\\
 =&\frac{T^2L}{\pi \ln(a_1 T)}\int_0^{\infty}\frac{u}{e^{u}-1}\int_0^1 \frac{\omega}{\sqrt{1-\omega^2}}d\omega du\\=& \frac{\pi T^2L}{ 6\ln (a_1T)}.
\end{split}
\end{equation}This is of order $T^2/\ln T$. On the other hand, if the smaller cylinder is Neumann, then the leading contribution to the thermal correction comes from the term with $n=0$ and $n=1$, which give
\begin{equation}
\begin{split}
\Delta_TE_{\text{Cas}}\sim  &\frac{L}{4\pi a_1^2}\int_0^{\infty}\int_0^{\infty}
 \frac{\omega^2}{e^{ \frac{\sqrt{\omega^2+k^2}}{aT}}-1}\frac{\omega dkd\omega}{ \sqrt{\omega^2+k^2}}
 \\=&\frac{L}{4\pi a_1^2}\int_0^{\infty}\int_{\omega}^{\infty}\frac{\omega^2}{e^{\frac{u}{aT}}-1}\frac{du}{\sqrt{u^2-\omega^2}}\omega d\omega\\
 =&\frac{L}{4\pi a_1^2}\int_0^{\infty}\frac{1}{e^{\frac{u}{aT}}-1}\int_0^u\frac{\omega^3}{\sqrt{u^2-\omega^2}}d\omega du \\
 =& \frac{ \pi^3LT}{90}a_1^2T^4.
\end{split}
\end{equation}This is of order $T^4$. Therefore, for electromagnetic cylinders with either perfectly conducting or infinitely permeable conditions, the leading term of the thermal correction is given by \eqref{eq6_30_1}.

\begin{acknowledgments}
 This project is funded by the Ministry of Higher Education of Malaysia   under the FRGS grant FRGS/2/2010/SG/UNIM/02/2.
\end{acknowledgments}

\end{document}